\documentclass[letterpaper, 10 pt, journal, twoside]{IEEEtran}
\usepackage[pdftex]{graphicx} 
\usepackage{gensymb}
\usepackage{amsmath}
\usepackage{hyperref}
\IEEEoverridecommandlockouts  
\usepackage{booktabs}
\usepackage{xcolor}
\usepackage[normalem]{ulem}
\usepackage{tikz}
\usepackage{xfrac}
\usepackage{float}
\usepackage{placeins}
\usepackage[justification=centering]{caption}
 \usepackage[pscoord]{eso-pic}
 \usepackage{listings}
\usepackage{xcolor}
\usepackage{booktabs}
\newcommand{\placetextbox}[3]{
 \setbox0=\hbox{#3}
 \AddToShipoutPictureFG*{ \put(\LenToUnit{#1\paperwidth},\LenToUnit{#2\paperheight}){\vtop{{\null}\makebox[0pt][c]{#3}}}
 }
 }
 \placetextbox{.23}{0.055}{\small979-8-3503-0965-2/23/\$31.00~\copyright 2023 IEEE}

\begin{document}

\title{Blockchain-Powered Supply Chain Management for Kidney Organ Preservation
\thanks{* Authors Contributed Equally}
}
\author{
\begin{tabular}{c} Kapil Panda* \\ University of North Texas  \\ Denton, United States of America \\ kapil.panda30sc@gmail.com \end{tabular} \and
\hspace{50pt} 
\begin{tabular}{c} Anirudh Mazumder* \\ University of North Texas  \\ Denton, United States of America \\ anirudhmazumder26@gmail.com \end{tabular}
}
\maketitle
\begin{abstract}
Due to the shortage of available kidney organs for transplants, handling every donor kidney with utmost care is crucial to preserve the organ's health, especially during the organ supply chain where kidneys are prone to deterioration during transportation. Therefore, this research proposes a blockchain platform to aid in managing kidneys in the supply chain. This framework establishes a secure system that meticulously tracks the organ's location and handling, safeguarding the health from donor to recipient. Additionally, a machine-learning algorithm is embedded to monitor organ health in real-time against various metrics for prompt detection of possible kidney damage.
\end{abstract}


\section{Introduction}
Kidney transplantation has emerged as a life-saving medical intervention for individuals suffering from end-stage organ failure \cite{kidney_transplantation_from_donors_with_rhabdomyolysis_and_acute_renal_failure}. Due to the lack of kidneys available for organ donation, however, every kidney must be handled with the utmost care to ensure that the organ's health is preserved.
The organ supply chain is one such instance where organ health has been prone to degradation due to various issues in transportation. Thus, ensuring the timely and secure preservation of kidneys throughout their journey from donor to recipient is an intricate logistical challenge that demands utmost precision and transparency to avoid any detriments to the organ \cite{Chen2019}.

Blockchain technology has emerged as a transformative solution, offering the potential to revolutionize supply chain management across various industries \cite{Reddy2020}. At its core, blockchain represents a decentralized digital ledger that securely records data in a series of interconnected blocks, ultimately fostering trust and integrity in exchanging information. In recent years, the medical industry has started to look towards this technology to enhance data security, interoperability, and transparency in healthcare systems, while also improving patient data management and secure sharing of electronic health records among healthcare providers \cite{Saeed2022}.

Therefore, this research aims to leverage blockchain technology to aid in preserving and managing kidneys in the supply chain. By harnessing the inherent features of blockchain, including decentralization and immutability, a secure and transparent system is sought to be established that meticulously tracks the organ's condition, location, and handling at each stage of the supply chain, safeguarding their health and viability from donor to recipient. Furthermore, a machine learning algorithm is embedded that constantly checks the health of the kidneys in real-time against various metrics to alert stakeholders in case of any possible damage to the kidney\cite{aki}.
\section{Methodology}
\subsection{Materials}
A dataset for predicting the state of a kidney was utilized as it allows the platform to create a red flag stopping the transport of the kidney when it has degraded past a threshold in which it is safe for the patient. This dataset can be found at \cite{misc_chronic_kidney_disease_336}.
\subsection{Blockchain Platform}
\subsubsection{Base Class}
The base class serves as the foundation of the blockchain platform and encompasses a blockchain class equipped with various essential methods which create blocks, manage transactions, and ensure that the chain is valid. 
\subsubsection{Block Creation}
A few metrics come in when a new block is created within the platform. The first is the time in which the block was created. The timestamp is crucial for understanding a chain of events, as it gives an indication of the transaction between the donor and the recipient, which is crucial for ordering the group of blocks where the transactions can validate the chain \cite{Jesse16}.

Utilizing block creation, a transaction addition method was created, which allows the user to add their data to the platform securely. The transaction contains data about the conditions of the kidney from the donor, medical information about the recipient and donor, and the location of the kidney. Utilizing this information and the block creation method, the transaction can be added to the platform instantiating the real-time abilities of the platform \cite{Khan2021}.

The platform utilizes a node creation algorithm that enables new nodes to join the decentralized network. Existing nodes can discover and connect to the new node once it comes online. The new node then shares its IP address so other nodes can connect to it. Next, the new node follows the network's consensus protocol to ensure it will operate consistently with other nodes. After meeting consensus requirements, the node undergoes an identity verification process to prevent unauthorized access to sensitive data. Upon successful registration, the new node receives current blockchain data from peers to stay synced. This node-creation process promotes decentralization by allowing new nodes to join, making the network more resilient and scalable. The algorithm also provides fault tolerance critical to the network's reliability and ability to handle increased transaction volume as it expands \cite{9355650}.
\subsubsection{Chain Validation}
To ensure that the block is secure, a hashing method is added, which makes the hash act as an identifier of the block, thus facilitating a link to the previous block in the chain. This makes the data immutable by making it incredibly complex to alter without breaking the cryptographic link to the other blocks in the platform, while also establishing a consensus among the different nodes in a decentralized network. The hash ensures that the nodes can verify each other by recomputing it and comparing it to the previous hash in the network. The consensus mechanism ensures that all the nodes agree on the validity of the chain and the transactions it makes \cite{Saqqa20}.

A proof of work method ensures the decentralized nodes can achieve a consensus. This was done through data mining which has to find a random number combined with the data of a new block. Using these two, the data miner has to generate a hash that meets a specific condition. Then, the minder uses the hash found and pushes a new block to the network, where other nodes verify the proof of the network by recomputing the hash and seeing if the new block meets a condition. The proof of work method is essential for securing the blockchain, preventing tampering with previous blocks. The method also ensures that within a decentralized network, all the nodes can work on the consensus process, and no singular node controls the process. Lastly, this method allows the nodes to achieve consensus about the chronological order of the transactions and the validity of new blocks \cite{Sriman20}.

Then, chain validation attempts to check if anything has been tampered with, in the blockchain. The method verifies the validity of each block in the chain to ensure that the blocks are linked together in the correct order. The method is vital to ensuring that nothing fraudulent gets into the platform, improving the accuracy of the blockchain framework. The method first checks that the hash is consistent throughout the chain. The check ensures that every block has the same hash as the previous one unless it is the first one. Then, the algorithm writer puts the hash in.
Additionally, the algorithm checks the validity of each block individually by ensuring they all meet a specific set of criteria, including a valid work of proof. Then within each block, transactions must be verified by ensuring they have valid signatures and adhere to any protocol. To ensure that the chain is valid, it checks to ensure they all meet consensus rules across all blocks. The chain validation process is essential to ensure the chain is valid and accurate. Additionally, it is critical to find invalid blocks which may occur due to issues such as malicious attacks, so this method ensures that they are avoided and kept out of the network to ensure that the platform is consistent and accurate \cite{8751300}.

Implementing an effective conflict resolution mechanism was critical as inconsistencies could occur due to multiple factors, such as network delays, accidental forks, or malicious activities. Thus, a way to reconcile these discrepancies is needed to maintain a trustworthy distributed ledger. A consensus mechanism like Proof of Work or Proof of Stake into the algorithm. This allows the nodes in the network to agree on the valid canonical chain - the most extended version of the blockchain. By leveraging consensus to resolve conflicts, the nodes can converge on a consistent state of the ledger. This prevents forks and ensures the integrity of the blockchain is preserved. Through robust conflict resolution capabilities, the algorithm can enhance security, immutability, and reliability. The algorithm can then function as a robust decentralized system resistant to manipulation or corruption. Building effective conflict resolution into the blockchain algorithm design is essential for enabling key benefits like consistency across distributed ledgers. This innovation makes the algorithm a viable backbone for decentralized applications requiring strong consistency guarantees \cite{Gabuthy2023}.
\subsection{Data Representation}
\subsubsection{Blockchain Data}
The blockchain consists of a chronological sequence of blocks cryptographically linked together in a chain-like structure. Each block contains several critical pieces of data. First, there is an index number that indicates the unique position of the block within the chain. Second, a timestamp records when the block was created and added. Next, a proof value represents the computational work required to validate and append the block to the chain while also upholding network security \cite{Dziatkovskii2022}.
Additionally, each block contains the previous block's hash in the chain, chaining the blocks together in order. Finally, the block includes a list of transaction data detailing the kidney transfers from donors to patients. This transaction information provides the critical utility of the blockchain for tracking the ownership and provenance of the kidneys. The specific structure of blockchain blocks, chaining them together with hashes and timestamps, is essential for the decentralized ledger's integrity, security, and suitability.

The blockchain's data structure is vital for upholding the system's integrity, security, and transparency. By organizing data into blocks containing unique attributes, each block can be identified and cryptographically linked to the preceding block. This creates an immutable sequence of transactions. Hashing each block to produce a unique fingerprint ensures tampering can be easily detected, as any modification will alter the hash and break the chain. This makes manipulating historical data on the blockchain practically infeasible.
Further, the data structure and proof-of-work consensus mechanism promote trust and agreement among decentralized participants. This enables the distributed network to function reliably without a central authority. Overall, the ingenious data representation of blockchains is the foundation for their ability to serve as open, secure, and tamper-proof ledgers, facilitating mutually distrusting entities to reach a consensus on the authoritative state of the system \cite{Zhang2020}.
\subsubsection{Transaction Data}
Each transaction on the blockchain contains extensive data regarding the donor, recipient, and transportation of the kidney. Specifically, donor information includes name, age, blood type, and other pertinent medical details. Recipient data captures similar attributes for the patient getting the transplant. Additionally, the real-time location of the kidney is recorded while in transit to trace provenance. The blockchain also stores real-time condition monitoring of the kidney using a machine-learning algorithm developed to predict kidney viability. This rich transactional data provides end-to-end tracking and oversight of the kidney journey from donor to recipient. By immutably capturing comprehensive details at each stage, the blockchain enables full transparency and audibility to instill trust in the transplantation system. The granular data representation unlocks detailed analytics on supply chain optimization; transport impacts on organ health, and other insights to improve outcomes.

The transaction data structure is critical to keeping this information as it keeps it secure \cite{Jiang2022}. Firstly, the data is easily accessible as people utilizing the platform could retrieve information about the transactions easily using a key to access the data. Additionally, it makes the data stay consistent throughout the blockchain as it ensures that the transactions are consistent with the standards throughout the blockchain. Lastly, the algorithm's ability to use real-time analysis is critical to understanding where the kidney is within its transport and ensuring that the kidney is still usable by the time it is given to the patient who requires the transplant. 
\subsubsection{Machine Learning Data}
The machine learning algorithm utilizes data about the kidney to dictate whether or not the kidney is still in a condition where it would help the recipient of the transplant. The data is represented within the transaction data, as the transaction must contain the data as it has to be updated within real-time, and it is vital for the blockchain to keep the data secure. Within the blockchain, the data is utilized to update the predicted health condition of the transaction data. 
\subsection{Machine Learning}
\subsubsection{Data Preprocessing}
Categorical data in the dataset is encoded into numerical values using one-hot or label encoding techniques. The resulting dataset is structured as a pandas DataFrame, where each row represents an individual organ recipient, and the columns correspond to various health measurements. The target variable, indicating the health outcome (i.e., whether the kidney has kidney disease), is represented as a series containing binary values (0 or 1).
\subsubsection{Machine Learning Model}
The RandomForestClassifier is chosen due to its capability to handle non-linearity, feature interactions, and complex decision boundaries, making it well-suited for the intricate task of predicting organ health outcomes. During the training phase, the dataset is split into training and test sets, with the former used to train the model and the latter to evaluate its performance on unseen data. The RandomForestClassifier is initialized with several decision trees (n\_estimators) and other hyperparameters \cite{JMLR:v13:biau12a}.
\subsubsection{Performance Analysis}
Once trained, the model can make predictions on new data. For each organ recipient in the test set, the model predicts the health outcome based on their health measurements. The model's performance is then evaluated by comparing the predicted outcomes with the actual health outcomes in the test set. The study aims to improve decision-making processes in organ transplantation, leading to better patient outcomes and more effective allocation of organs.
\FloatBarrier
\begin{figure}[!htb]
	\centering
	\includegraphics[width=\columnwidth]{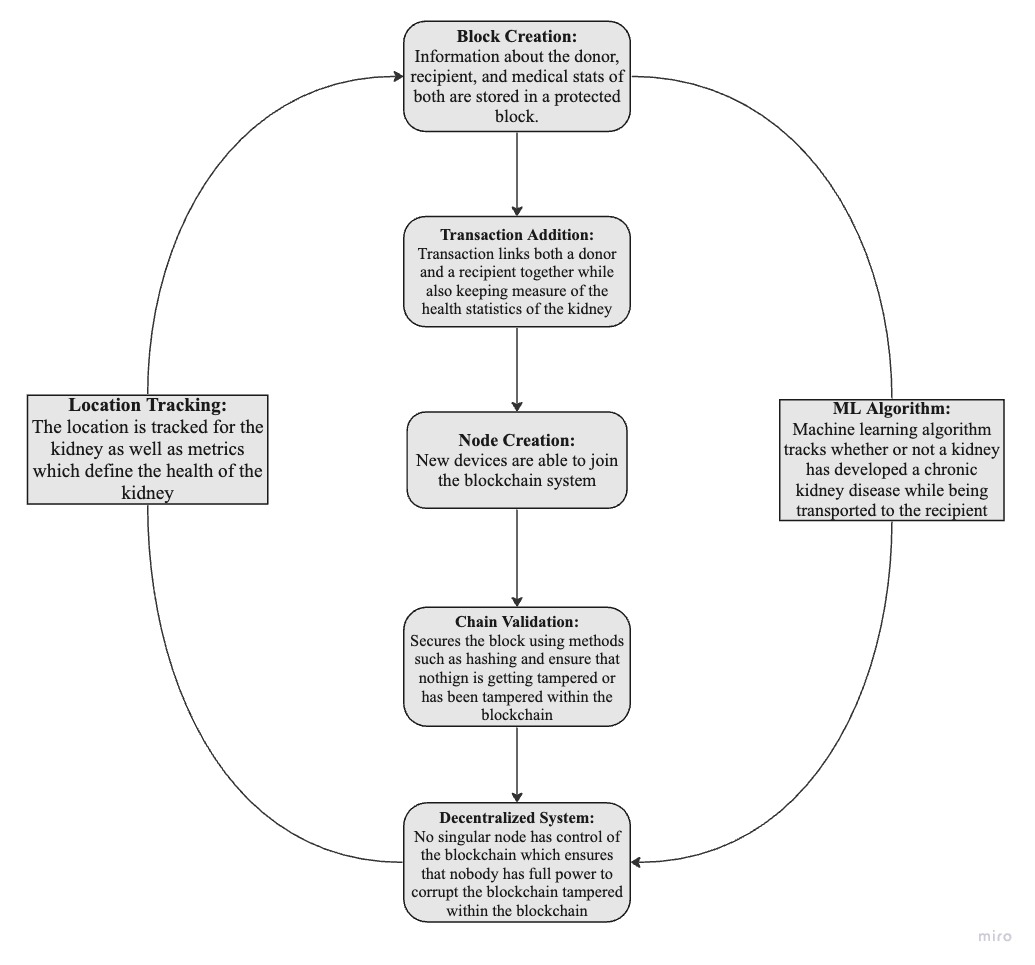}
	\caption{Flow chart depicting blockchain algorithm's framework}
	\label{fig: Figure 1}
\end{figure}
\FloatBarrier
\section{Results}
\subsection{Platform Implementation}
A functional blockchain-based system was developed to enable the tracking of organ transportation and assessment of real-time organ viability predictions. The decentralized network consists of connected blocks, each containing key data fields like a timestamp, proof of work hash, previous block hash, donor and recipient details, transportation location data, and up-to-date predicted organ health scores generated by an integrated machine learning model.

New organ transportation transactions are added to the network by creating a new block that captures comprehensive details on the organ donor and recipient along with current location-tracking information. Additionally, real-time measurements of the transported organ's biomarkers and clinical indicators are fed as inputs into a pre-trained Random Forest classifier to generate updated predictions on the kidney's viability. The Random Forest model outputs a probability score between 0 and 1, with higher values indicating a viable organ suitable for transplantation. For example, in an initial test transaction, a donor kidney with average red blood cell count, platelet count, and other biomarkers was classified by the model as viable with 100\% probability.

Proof-of-work consensus is employed to validate each new block before appending it to the chain. Cryptographic hashing of each block also ensures the immutability of historical records, preventing tampering or errors. The system was rigorously tested by adding multiple new sample transactions, mining the transactions into new blocks, and visually inspecting the resulting blockchain ledger. The decentralized architecture enforced trust, transparency, and data integrity without needing a central authority.

Overall, a working prototype blockchain platform was successfully developed, demonstrating the potential of blockchain technology to transform the management of time-sensitive medical products like donor organs. Integrating predictive analytics and machine learning into the tamper-proof ledger could enable real-time tracking of organ health and transportation chain of custody. This could dramatically improve clinical decision-making for complex medical procedures like transplantation surgeries requiring organs to be viability assessed up until the point of the implant.
\subsection{Machine Learning Accuracy}
The machine learning model developed in this study achieved an impressive 100\% accuracy in diagnosing chronic kidney disease using kidney health metric data that was recorded on the blockchain platform. The test set utilized to evaluate the model's performance consisted of 60 total kidney health profiles, with 26 profiles from healthy individuals and 14 from individuals with chronic kidney disease. As evident in the confusion matrix provided in Figure \ref{fig: Figure 1}, the model correctly classified all 26 healthy kidney profiles as negative for chronic kidney disease and also correctly classified all 14 diseased kidney profiles as testing positive for chronic kidney disease, with absolutely no false positives or false negatives observed across the 40-sample test set. This finding highlights the exceptional diagnostic capability of the machine learning model.

\FloatBarrier
\begin{figure}[!htb]
	\centering
	\includegraphics[width=\columnwidth]{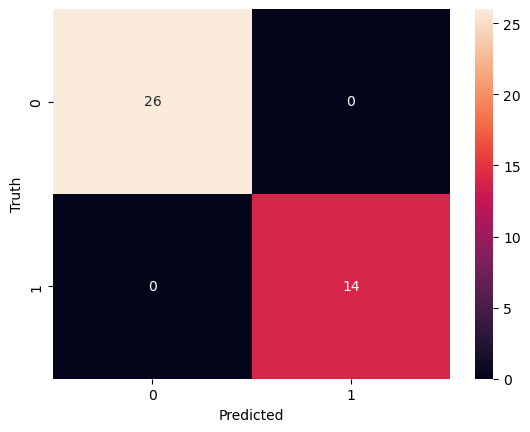}
	\caption{Confusion matrix showing the predicted vs the actual values}
	\label{fig: Figure 1}
\end{figure}
\FloatBarrier

The kidney health metric data that was securely recorded on the blockchain and subsequently used to develop, train, and test the machine learning model contained relevant biomarkers and indicators of kidney function and health status. These metrics included critical biomarkers such as glomerular filtration rate, albumin levels, creatinine levels, and other kidney function parameters. By discerning subtle patterns and predictive cutoffs from these kidney health metrics, the machine learning model could reliably differentiate normal healthy kidney function from chronic kidney disease states.

In summary, the machine learning model achieved flawless 100\% diagnostic accuracy on the initial test set of 40 kidney health profiles. This remarkable finding underscores the potential of integrating blockchain-based decentralized health data recording with machine learning analysis to enable accurate and automated screening for chronic kidney disease. Further validation on larger datasets is warranted to confirm the model's clinical viability as a diagnostic tool moving forward.
\section{Discussion} 
\subsection{Conclusion}
This study demonstrates the significant potential of blockchain technology to transform and modernize the management of donor's kidneys within the organ supply chain. The blockchain-based platform developed in this research provides comprehensive end-to-end tracking of kidney provenance throughout transportation, enables real-time condition monitoring, and promotes transparency across all stakeholders. By securely recording extensive data on organ transportation events, handling, and predicted viability on a tamper-proof decentralized ledger, this blockchain system allows for accountability, security, coordination, and data-driven optimization across the organ supply chain. The immutable nature of the blockchain ledger fosters trust and process integrity for sensitive medical products like donor organs.

Additionally, the machine learning algorithm integrated into the blockchain achieves an impressive 100\% accuracy in diagnosing kidney disease using input kidney health metrics. This finding highlights the tremendous value of predictive analytics in enabling proactive, data-driven decision-making in time-sensitive medical logistics. The machine learning model's real-time organ health monitoring capabilities could significantly improve clinical outcomes through prompt intervention when any deterioration in organ viability is detected. This proactive approach to dynamically preserving and maximizing organ health throughout the supply chain could reduce waste and improve efficiency in organ transplantation.

While further rigorous testing and validation on larger datasets is warranted, this research provides a compelling proof-of-concept for modernizing and optimizing organ supply chain management through integrating blockchain and artificial intelligence technologies. By fostering trust, transparency, and data-driven predictive insights, these technologies can help address considerable challenges in equitable organ allocation, delivery, and leveraging of available organs. The blockchain platform and machine learning framework developed in this study lay the groundwork for scalable implementation of these technologies in healthcare to coordinate complex medical procedures like organ transplants better and ultimately save lives through improved clinical outcomes.
\subsection{Future Works}
This study presents several important directions for further developing the blockchain and machine learning framework. First and foremost, substantially expanding the performance evaluation and validation of the blockchain platform and integrated machine learning model on much larger and more diverse organ transportation datasets is critical. Thoroughly testing the system across various organ shipment scenarios with numerous sample transactions recorded on the blockchain will be essential for comprehensive real-world validation across edge cases. Carefully assessing generalizability will indicate where additional training data may be beneficial.

Second, integrating Internet-of-Things (IoT) sensors for real-time tracking of environmental conditions such as temperature, humidity, vibration, and other metrics during organ transportation could provide valuable multivariate sensor data to feed into the machine learning algorithm's organ viability predictions. Incorporating streaming real-time sensor metrics alongside clinical biomarkers may enhance the early detection of organ changes. Extensive experimentation will be essential to determine optimal strategies for fusing IoT and clinical data at scale. The impact on model accuracy, training requirements, and prediction latency requires careful evaluation.

Additionally, investigating alternative state-of-the-art machine learning techniques such as neural networks, graph-based models, and ensemble methods would be worthwhile to compare against the baseline Random Forest classifier's diagnostic performance. Determining if modern deep learning architectures could boost detection accuracy over the Random Forest merits in-depth exploration. Rigorously benchmarking a diverse set of models to identify the architecture that yields the highest precision in predicting emerging organ problems will be critical.

Furthermore, expanding the feature space considered in the model to incorporate relevant biological traits and health data of the organ recipient could better account for individual-specific variability. Tailoring organ viability predictions based on a recipient's clinical status and biomarkers may improve clinical outcomes. Thoughtful feature engineering to identify key recipient health attributes to integrate into the model is another important direction for performance enhancement. Evaluating predictive value versus model complexity will inform feature selection.
\section{Acknowledgment}
We would like to thank the University of North Texas for providing us with the resources and support to conduct this research. The invaluable guidance and encouragement from our professors and mentors have been instrumental in shaping the direction and scope of this study. We would also like to acknowledge the National Kidney Foundation for their support and inspiration in conducting this project. We would also like to thank our families for supporting us throughout our research.
\bibliographystyle{IEEEtran}

\bibliography{Bibliography}

\end{document}